**Title:** Contaminant detection using a CZT photon counting detector with TDI image reconstruction
**Authors:** Joanna Nguyen[1], Devon Richtsmeier[1], Kris Iniewski[2] and Magdalena Bazalova-Carter[1]
[1]Department of Physics and Astronomy, University of Victoria, 3800 Finnerty Road, Victoria, Canada, V8P 5C2
[2]Redlen Technologies, 1763 Sean Heights, Saanichton, Canada, V8M 1X6

**Abstract**
Food x-ray inspection systems are designed to detect unwanted physical contaminants in packaged food to maintain a high level of food safety for consumers. Modern day x-ray inspection systems often utilize line scan sensors to detect these physical contaminants but are limited to single or dual energies. However, by using a photon counting detector (PCD), a new generation of food inspection systems capable of acquiring images at more than two energy bins could improve discrimination between low density contaminants. In this work, five type of contaminants were embedded in an acrylic phantom and imaged using a cadmium zinc telluride (CZT) PCD with a pixel pitch of 330 μm. A set of images were acquired while the phantom was stationary, and another set of images were acquired while the phantom was moving to mimic the movement of a conveyor belt. Image quality was assessed by evaluating the contrast-to-noise ratio (CNR) for each set of images. For imaging times larger than 25 ms, the results showed that the moving phantom data set yielded larger CNR values compared to a stationary phantom. While conventional x-ray inspections often utilize line scan sensors, we report that physical contaminant detection is possible with a CZT PCD x-ray imaging system.

**1. Introduction**
The use of photon-counting detectors (PCDs) in medical imaging systems has become increasingly common in the past few years given their superior spatial resolution, multi-energy bin capabilities and improved image quality compared to conventional energy-integrating detectors (Leng et al 2019). While these PCD systems are still currently under research and development, studies have shown promising results in clinical applications such as breast, temporal bone and chest computed tomography (CT) among other potential medical applications (Kalluri et al 2013; Leng et al 2016; Barlett et al 2019; Leng et al 2019). PCDs are advantageous in that they can directly convert the incident photon into a cloud of charges by interacting with the detector material, which results in an electronic signal that is proportional to the energy of the incident photon (Willemink et al 2018). While the detector material can vary widely, materials such as germanium, gallium arsenide, cadmium telluride and cadmium zinc telluride (CZT) are commonly used, and the advantages and disadvantages of each material are further discussed elsewhere (Pennicard et al 2017). Given the multi-energy bin capabilities of PCDs, electronic noise can be easily discarded by setting an energy threshold above the level of noise. By setting additional energy thresholds, incoming photons can be sorted into different energy bins, creating multiple energy selective images from a single exposure. This is highly advantageous given that low density materials that are typically difficult to distinguish in conventional x-ray imaging with energy-integrating detectors, would be better discriminated in these low energy bins.

Aside from the medical industry, PCD-based inspection systems have been developed for the food inspection industry to detect physical contaminants in packaged food products, though literature data on these proprietary systems are limited. Physical contaminants from the environment and processing line can cause glass, stones, metal, wood, plastic and other foreign objects to be integrated into the final food product, which can be a choking hazard and cause harm to the consumer (Haff and Toyofuku 2008). Therefore, it is crucial to ensure these contaminants are detected prior to shipping as the impact of large-scale food recalls can be significantly detrimental to company profit margins and the damage to a company's reputation may have a lasting effect.

Given the high conveyor belt speeds used in the food industry, previous iterations of food inspection imaging systems typically used line scan image detectors (Haff and Toyofuku 2008). These detectors contain a single row of image sensors that allows a 2-dimensional (2D) image to be continuously constructed line-by-line as the object passes through the sensor. Traditionally, these line scan sensors were limited to a single energy, though modern inspection systems can employ dual-energy line sensors, which can improve discrimination between various materials. In contrast, area scan image sensors are better suited for stationary objects as it can capture a 2D image and are generally not suited for moving objects. Since PCDs can acquire images at more than two energy bins, discriminating between various types of materials would be beneficial and easier to achieve. However, using an area PCD would require the use of a time-delay and integration (TDI) algorithm to reconstruct the final 2D output image, especially if the object does not fit in the field-of-view of the detector (Dörr et al 2012). For these types of measurements, charge-coupled devices are encoded with the TDI mode and the charge generated in each detector element row is shifted in synchrony with the object being imaged (Wong et al 1992). For a detector with M rows of pixels, the charge generated in rows M, M+1 and M+2 etc. are shifted (time-delay) and summed (integration) by applying a potential across each row and the summed signal represents the final output (Semwa and Saxena 2020). However, in this work, the principles of TDI were applied to an in-house algorithm for post-processing image reconstruction.

Previously, our research group developed an x-ray imaging system that utilized a CZT PCD to detect several contaminants embedded in a phantom that remained stationary during the image acquisition period (Richtsmeier et al 2021). In this work, we present an x-ray imaging system with a large area CZT PCD capable of detecting moving contaminants at a constant speed of 1 cm/s, 2 cm/s and 4 cm/s during image acquisition to mimic a conveyor belt.

## 2. Materials and Methods
2.1 Experimental set-up
The CZT PCD (Redlen Technologies, Saanichton, BC, CA) was connected to a linear motion stage and mounted vertically on an optical table (Newport Corporation, Irvine, CA, USA) to allow the PCD to move perpendicular to the table. The PCD consisted of two detector modules which had 24×576 pixels in total and the pixel pitch was 330 µm, resulting in a total detector size of 7.9×190.1 mm. At a distance of 50 cm from the PCD, a Comet MXR-160/22 x-ray tube (Comet Technologies, San Jose, CA, USA) was connected to an X-Y motion stage to allow movement along two planes as shown in Figure 1a. Between the PCD and x-ray tube, a sample stage was connected to a linear stage and attached horizontally to the optical table to allow for movement parallel to the table surface. The phantom used in this study was made of 2 cm thick acrylic with a height and width of 14 cm and 6 cm, respectively. Five different types of materials, which represent physical contaminants were embedded in an acrylic phantom and made flush with the surface of the phantom (Redlen Technologies, Saanichton, BC, CA) as shown in Figure 1b. The contaminants in Figure 1b from top to bottom correspond to polytetrafluoroethylene, which will be referred to as Teflon hereinafter, polypropylene, calcium carbonate, glass and stainless-steel. The first three materials are cylindrical rods of varying thicknesses while the glass and stainless-steel are beads of varying diameters, as summarized in Table 1. The phantom was fixed on the sample stage and positioned equidistant between the x-ray tube and detector at a distance of 25 cm for all measurements.

Figure 1: The a) experimental set up for all measurements where the CZT PCD is on the left, the x-ray tube on the right while the acrylic phantom is positioned equidistant between the CZT PCD and x-ray tube and b) the acrylic phantom embedded with various contaminant materials.

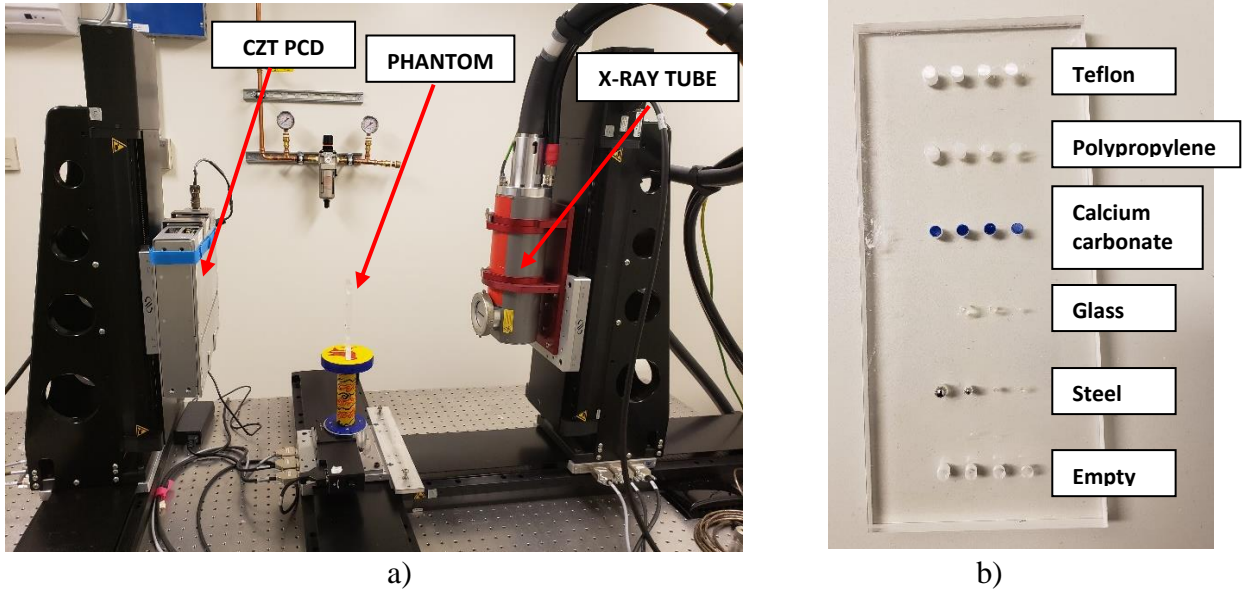

Table 1: The dimensions of the five contaminants embedded in the acrylic phantom.

|  | Contaminant Material | Diameter (mm) | Thickness (mm) |
|---|---|---|---|
| Cylindrical | Teflon<br>Polypropylene<br>Calcium carbonate | 3.0<br>3.0<br>3.0<br>3.0 | 2.0<br>4.0<br>6.0<br>8.0 |
| Spherical | Glass | 1.0<br>2.0<br>3.0 | -<br>-<br>- |
| Spherical | Stainless-steel | 0.3<br>0.5<br>1.0<br>2.0 | -<br>-<br>-<br>- |

For all measurements, the x-ray tube voltage was fixed at 120 kVp and the tube currents were 0.5 mA and 1 mA. The focal spot was 1 mm and measurements were taken with and without a 2 mm thick aluminum (Al) filter. A 0.7 cm thick lead collimator with a beam opening of 3 cm by 0.3 cm was used to create a broad beam which covered the entire detector area. The energy thresholds were chosen to be 16, 30, 50, 70, 90 and 120 keV, which resulted in five energy bins between these six energy thresholds. These thresholds correspond to the five energy bins shown in Table 2. A 6th energy bin was created to collect any pulse pileup effects from photon energies greater than 120 keV while a 7th energy bin was created as a total counts bin, which is the summation of energy bins 1-5. The minimum energy threshold of 16 keV was chosen to eliminate electronic noise.

Table 2: The corresponding energy range for each energy bin

| Energy Bin Number | Energy Range (keV) |
|---|---|
| 1 | 16-30 |
| 2 | 30-50 |
| 3 | 50-70 |
| 4 | 70-90 |
| 5 | 90-120 |
| 6 | >120 |
| 7 | 16-120 |

2.2 Data acquisition and image reconstruction

For all measurements, projection images of each contaminant were captured over a 200 ms acquisition time at a frame rate of 1000 fps. Two types of images were acquired, namely one where the phantom was stationary and another where the phantom was moving at a speed of 1 cm/s, 2 cm/s and 4 cm/s on the sample stage to mimic the movement of a conveyor belt. For the stationary phantom measurements, two sets of data were acquired at 0.5 mA with and without the 2 mm Al filter and another two data sets were acquired using a tube current of 1 mA with and without the 2 mm Al filter. An in-house algorithm was written in Python to capture each measurement and the resulting data was organized into the specified energy bins (Python Software Foundation, https://www.python.org/) while Matlab (The Mathworks, Natick, MA, USA) was used for all data analysis and image reconstruction.

2.2.1 Pixel correction

A single flat field and dark field image were acquired for 60 s in absence of the phantom. The flat field image was acquired while the x-ray beam was turned on while the dark field image was obtained with the x-ray beam turned off. Both the flat field and dark field images were scaled down to 1 ms equivalent images by normalizing the 60 s images. Scaling these images down to a 1 ms equivalent image resulted in less variations across the detector pixels compared to normalizing with 1 ms images. The flat field corrected image, $I_{CORR}$, was obtained by using Eq. 1:

$$I_{CORR} = -ln\left(\frac{I_P - I_{DF}}{I_{FF} - I_{DF}}\right) \qquad (Eq.1)$$

Where $I_P$ is the projection image of the phantom, $I_{DF}$ is the dark field image and $I_{FF}$ is the flat field image. To minimize over- and under-responsive pixels in each projection image, another 60 s flat field image was acquired and the natural log ratio of the two flat field images were calculated to find pixels that had variations greater than 5%. Nearest-neighbour interpolation was used to correct for these over- and under-responsive pixels.

2.2.2 CNR Calculations
The contrast-to-noise ratio (CNR) for each type of contaminant was evaluated for both the stationary and moving phantom data using Eq. 2:

$$CNR = \frac{\mu_{ROI} - \bar{\mu}_{BKG}}{\bar{\sigma}_{BKG}} \qquad (Eq.\,2)$$

Where $\mu_{ROI}$ is the mean signal in the region of interest (ROI) of each contaminant while $\bar{\mu}_{BKG}$ and $\bar{\sigma}_{BKG}$ are the mean and standard deviation of the phantom background ROI along the acrylic portion of the phantom and averaged across four different ROIs to further minimize pixel response variations. For the stationary phantom data sets, it was possible to fix the location of the background ROIs for all measurements. However, this was not possible for the moving phantom data set since the phantom moving at 2 cm/s and 4 cm/s travels across a larger distance across the detector compared to using a speed of 1 cm/s for the same number of projections. Therefore, averaging the background ROIs for the moving phantom data set is highly advantageous to minimize any pixel response variation. Based on the Rose Model, a CNR ranging between 3-5 is required for an object to be considered detectable and a CNR of 4 will be used as a baseline to assess the image quality of each contaminant material (Rose and Biberman 1975). The relative noise was also investigated and is calculated by taking the ratio between the standard deviation and mean of the phantom background averaged over the four ROIs.

2.2.3 Time delay and integration (TDI) image reconstruction
For measurements using the stationary phantom, the resulting projection images were cumulatively summed and the CNR was calculated for each cumulative sum for each contaminant material and size. However, for data resulting from the phantom moving at each of the three speeds, a time delay and integration (TDI) algorithm was applied prior to calculating the CNR. This TDI algorithm is necessary to align the contaminants and allow the projection images to be cumulatively summed to reconstruct the final image. Without this TDI algorithm, the contaminants would be smeared across the final image and their detection would be compromised as shown in Figure 2e.

For the phantom moving at 1 cm/s, it was observed that for every 18 projections, the contaminants shifted across one detector pixel length. Therefore, for the first 18 projections, no pixel shift was required, however the next group of 18 projections were shifted by one column of pixels opposite to the direction of motion. This pixel shift is performed by removing the first column of pixels in the image and stitching it to the end (right hand side) of the image. However, for the third group of 18 projections, the first two columns of pixels were removed and shifted to the end of the image and so forth. This was repeated until all 200 projections were shifted accordingly and each projection image was cumulatively summed. The same principles were applied to the 2 cm/s and 4 cm/s data sets, though the step size was reduced by a factor of two and four, respectively. The floor function in Matlab was used to round the step size in the 4 cm/s data set to the nearest integer that is equal or less than the resulting value, given that the step size was 4.5.

Results from imaging the stainless-steel contaminants moving at 4 cm/s are shown in Figure 2 where the left column (a-d) represents the pre-shifted images, and the right column (f-i) shows the post-shifted images for the 4 cm/s data set. Between Figures 2a and 2f, no shifting is required. However, for the 50th image, the first 10 columns are taken in Figure 2b and stitched to the right side of the image as demonstrated in Figure 2g. The number of pixel columns required to shift the contaminants in Figures 2c and 2d correspond to 21 and 52, respectively. After aligning the contaminants, the ROIs for the background and contaminants were created to calculate the CNR and the images were cumulatively summed. Figure 2e represents the output image when the projection images are cumulatively added without applying the TDI algorithm which resulted in a smeared image, while Figure 2j shows the output image after implementing the TDI algorithm.

Figure 2: A comparison between the flat field corrected (a-d) pre-shifted images and (f-i) post-shifted images using TDI image reconstruction for the stainless-steel contaminants moving at 4 cm/s. Summing the projection images e) without and j) with the TDI image reconstruction.

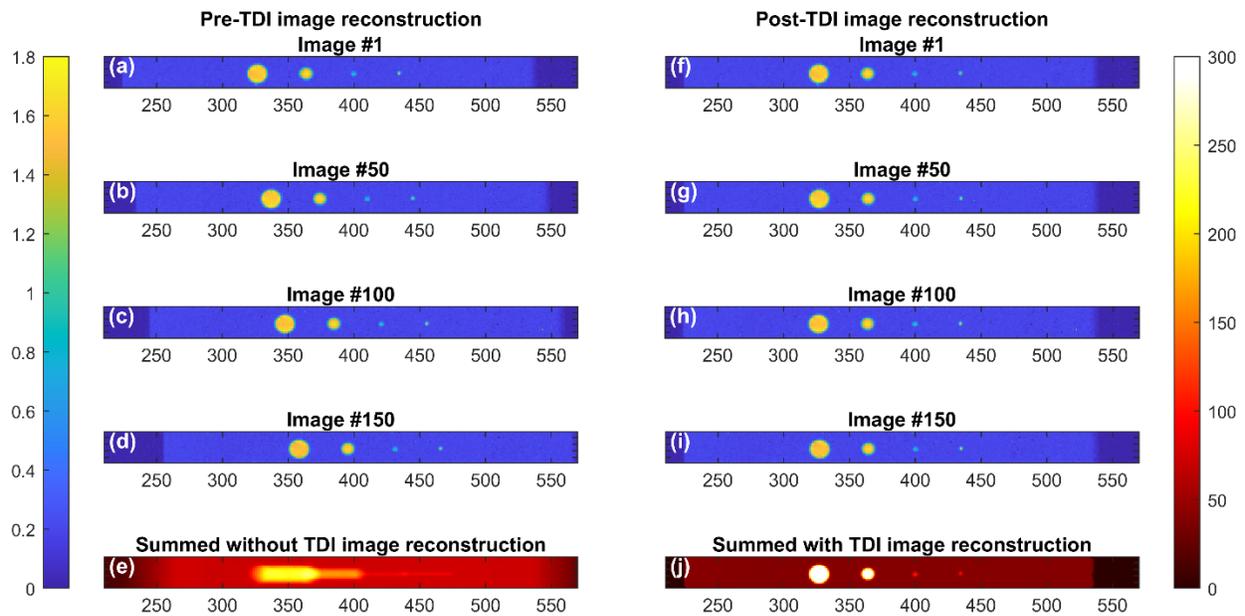

## 3. Results and Discussion
3.1 Stationary phantom data
3.1.1 Effects of beam filtration on CNR
In the presence of a 2 mm external Al filter, the CNR for bin 2 outperformed the other energy bins for acquisition times greater than 100 ms, as shown in Figure 3. For these measurements, the tube current was set to 1 mA and the tube voltage was 120 kVp, while the left and right columns in Figure 3 correspond to measurements in the absence and presence of the 2 mm Al filter, respectively.

In the presence of the Al filter, low energy x-rays will be largely absorbed by the filter. Given that energy bin 1 comprises of low energy photons between 16-30 keV, a 2 mm Al filter will attenuate 98% of a 16 keV photon beam and 45% of a 30 keV photon beam, which could explain why the CNR for all bins decreased in the presence of the filter with the exception of bin 2. The filter is also advantageous in filtering out low energy photons that would contribute to noise and degrade image quality. Another consequence of beam filtration is beam hardening, which leads to a larger average beam energy compared to a non-filtered beam. This energy shift in the beam could lead to a lower CNR given that the beam results in lower contrast.

In some cases, the bin 7 provides the largest CNR for acquisition times ranging between 1-20 ms, whereafter the CNR from bin 2 is dominant, thereby suggesting that an Al filter should be used in cases where long acquisition times are required. For bins 1 and 7, the CNR begins to plateau at 25 ms, which suggests that increasing the acquisition time beyond this time point would negligibly affect the CNR.

Figure 3: The CNR for the stainless-steel contaminants with diameters of 0.3 mm (a, e), 0.5 mm (b, f), 1 mm (c, g) and 2 mm (d, h) as a function of time for each energy bin. The tube current and voltage was 1 mA and 120 kVp, respectively. No external filtration was used in the left column (a-d) while a 2 mm Al filter was used in the right column (e-h).

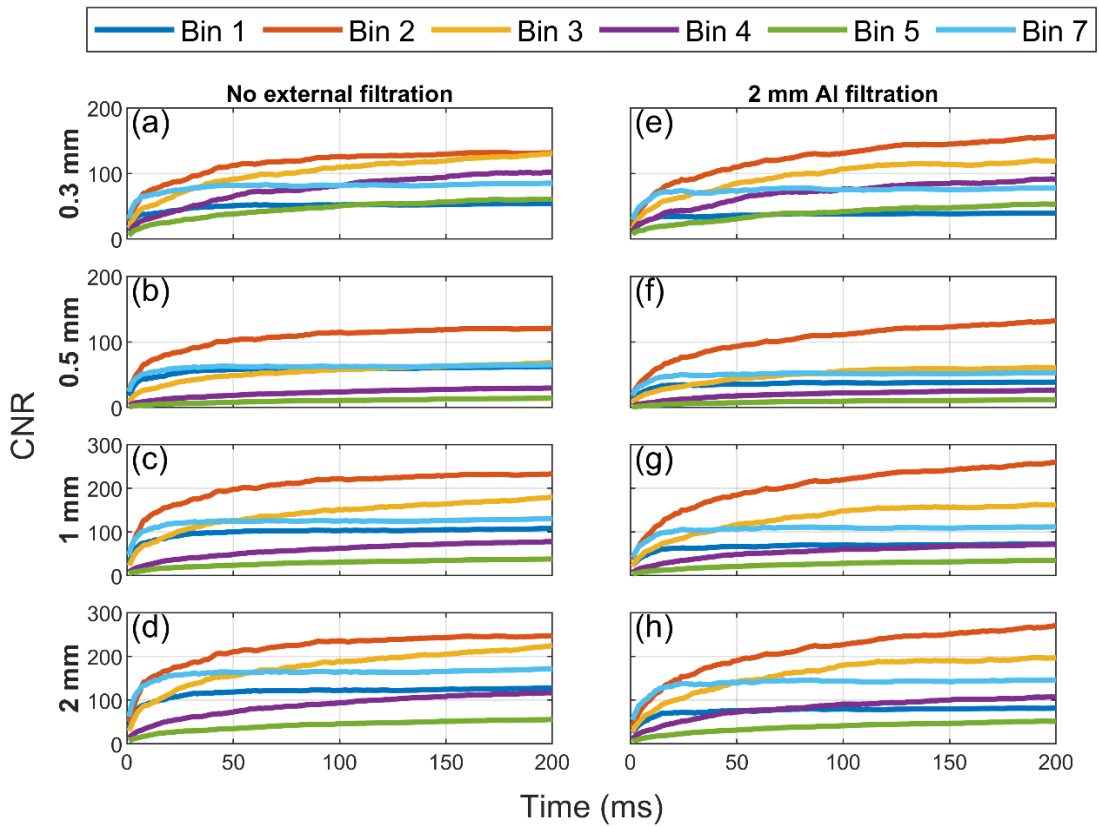

### 3.1.2 Effects of tube current on CNR

Using a tube current of 1 mA resulted in larger CNR values for all energy bins compared to a tube current of 0.5 mA, as shown in Figure 4 for the glass contaminants. With the exception of the 1 mm diameter glass bead at 0.5 mA, the CNR in bin 2 is much larger compared to the other bins for acquisition times longer than 25 ms. This can be explained by looking at the relative noise in Figure 5 in the phantom background, which is the smallest for bins 2 and 7. Since the noise plateaus at around 50 ms, increasing the acquisition time beyond 50 ms will not significantly improve the CNR.

The total number of detected photons from using the tube currents of 0.5 mA and 1 mA is shown in Figure 6a. By doubling the tube current, it is expected that the number of photons will increase by a factor of 2 as demonstrated by the dotted line in Figure 6b. Given that noise is inversely proportional to the square root of number of photons, doubling the tube current should reduce the noise by a factor of 1.4 as indicated by the dotted line in Figure 6c. Since bin 6 is designated as the pulse pile up bin, the number of counts in this bin makes up 0.3% of the total counts. As a result, taking the ratio of the number of photons did not agree with the expected ratio of 1.4. In addition, it is likely that the first energy threshold of 16 keV allowed some electronic noise to pass through, which also lead to a larger error for bin 1.

Figure 4: The CNR for the glass contaminants with diameters of 1 mm (a, d), 2mm (b, e) and 3 mm (c, f) as a function of time for all energy bins. The left column (a-c) and right column (d-f) correspond to measurements that used a tube current of 0.5 mA and 1 mA, respectively with no Al filtration.

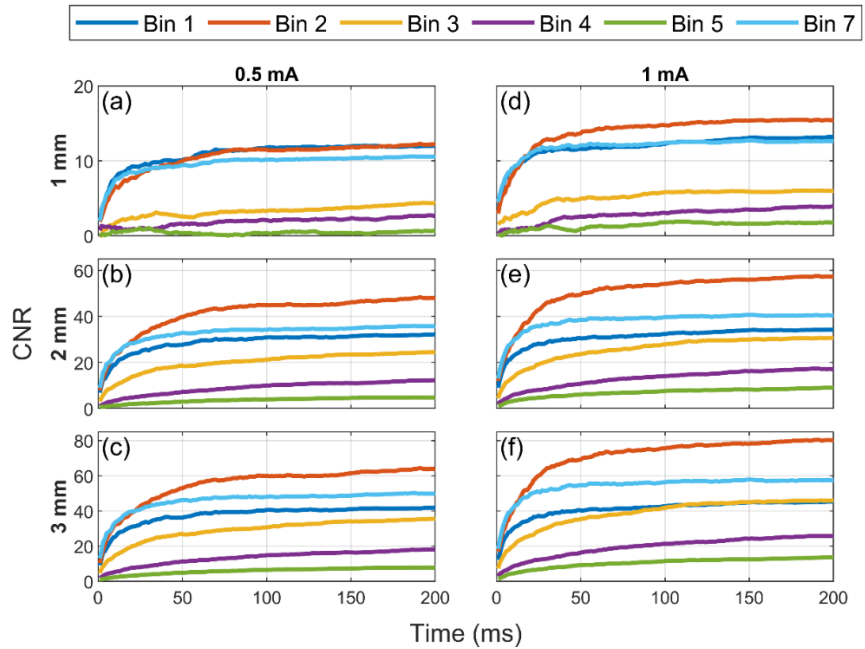

Figure 5. The relative noise in the phantom background for all energy bins when using a tube current of 0.5 mA and 1 mA with no additional filtration.

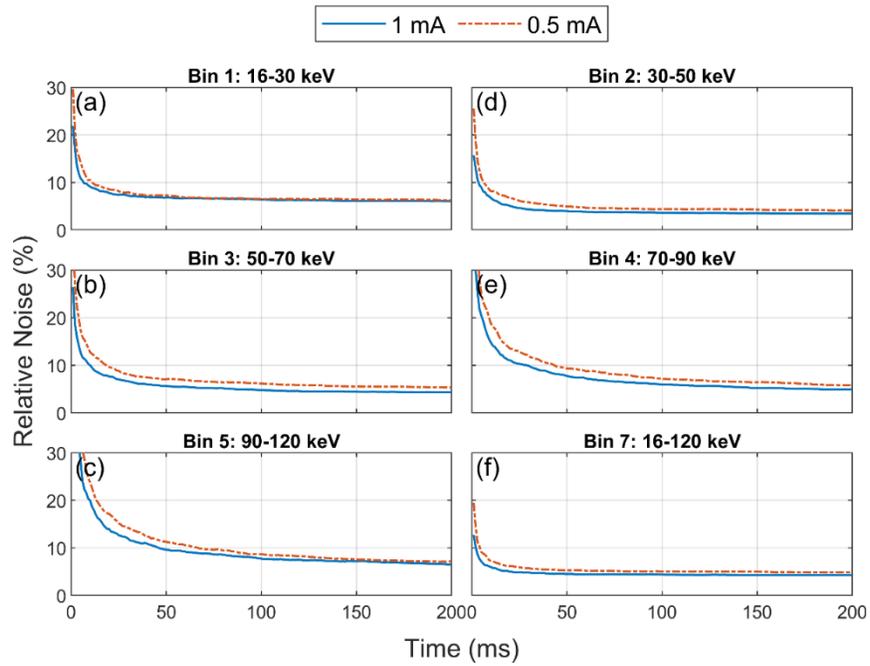

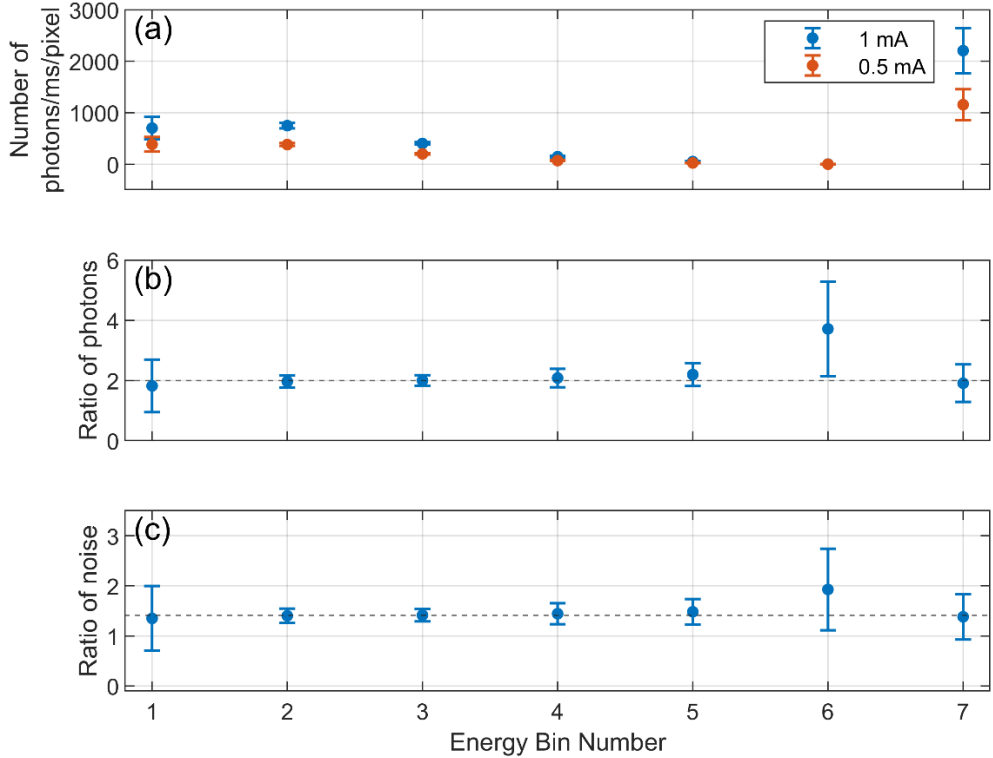

Figure 6: The a) number of photons/ms/pixel from two flat field scans at 0.5 mA and 1 mA with no external filtration, b) ratio of number of photons and c) the ratio of noise. It is expected that doubling the tube current would increase the number of photons by a factor of 2 and the noise would be reduced by a factor of 1.4, as demonstrated by the dotted lines.

3.1.3 Effects of contaminant thickness on CNR

Plotting the CNR as a function of cylindrical contaminant thickness for calcium carbonate (Figure 7 a-d), polypropylene (Figure 7 e-h) and Teflon (Figure 7 i-l), shows the CNR is the largest for bins 2, 3 and 7 for all thicknesses. Since these contaminants are less attenuating than stainless-steel and glass, it is expected that the CNR would be smaller. Based on the results for polypropylene, the CNR in bin 7 outperforms the other energy bins, though in the case of Teflon, the CNR in bin 2 exceeds the other energy bins. For the calcium carbonate contaminants, it was found that the material thickness was not homogenous across the contaminant ROI used to calculate the CNR, which could explain the inconsistent trends in the CNR.

With increasing material thickness, incident photons are exponentially attenuated over a larger distance, which resulted in a larger signal difference between the contaminant ROI ($\mu_{ROI}$) and background ROI ($\mu_{BKG}$), thereby leading to an increasing CNR. For the smallest material thickness of 2 mm, the signal difference between the contaminant and background ROIs is reduced, which makes it more difficult to distinguish thin and less dense objects from the phantom background. Overall, the CNR is the lowest for calcium carbonate and highest for Teflon, which is attributed to the differences in the linear attenuation coefficients compared to the acrylic background.

Based on the Rose criterion, the 2-6 mm thick calcium carbonate and 2-6 mm thick polypropylene contaminants could be considered undetectable as the CNRs for all energy bins are less than 4. For the 2 mm Teflon (Figure 7i) contaminant, only bin 2 yields a CNR greater than 4 after a time point of 25 ms, while the CNR remains below this baseline for the other energy bins. As the material thickness increases from 4-8 mm for Teflon (Figure 7j-l), the number of energy bins that yield a CNR greater than 4 also increases. Based on these results, significant improvements to the system will be required to better distinguish these contaminants that have a material thickness of less than 4 mm from the phantom background, which is made from acrylic.

Figure 7: The CNR for the calcium carbonate (a-d), polypropylene (e-h) and Teflon (i-l) with increasing material thickness for each column starting from the left. The tube current was 1 mA with 2 mm Al filtration.

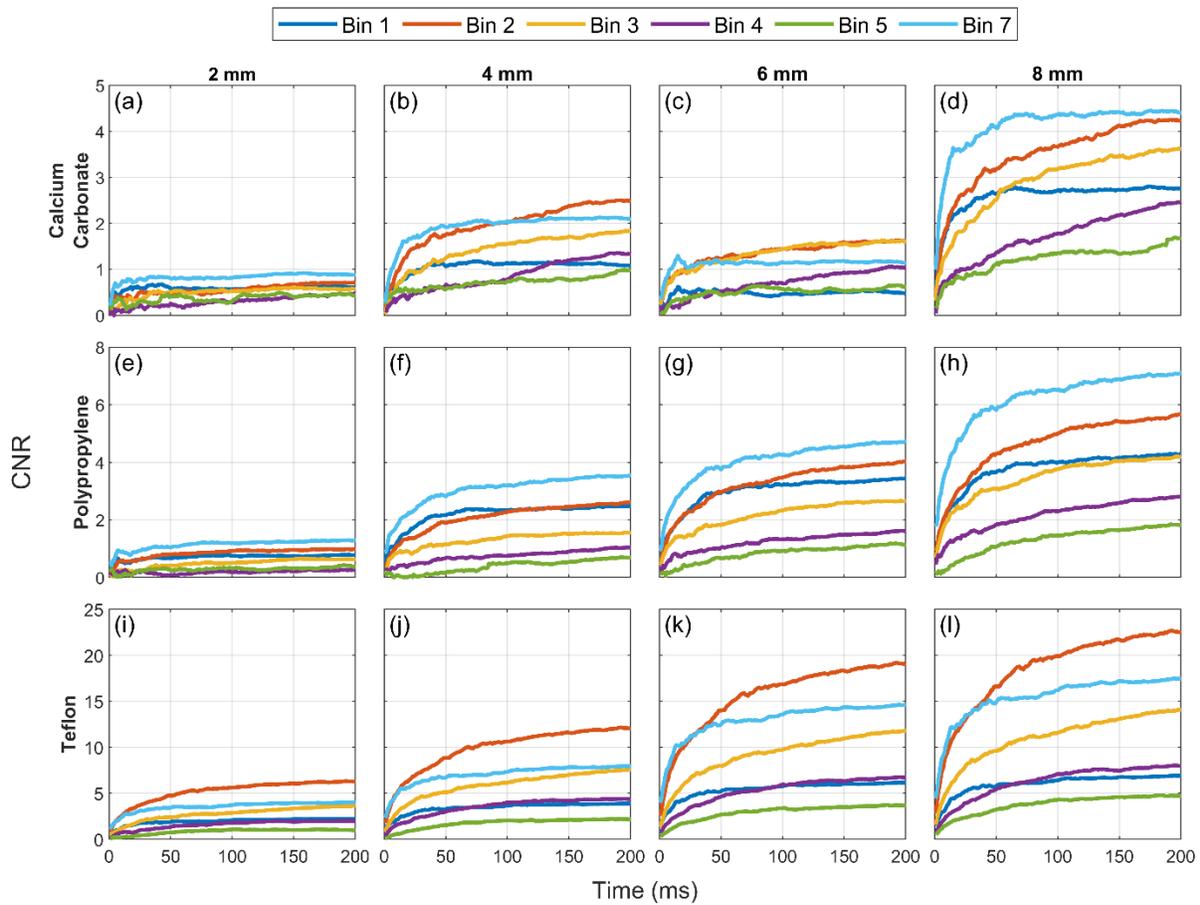

## 3.2 Moving phantom data

The CNR for the glass contaminants while the phantom is stationary, moving at a speed of 1, 2 and 4 cm/s is shown in Figure 8. As seen previously in Figure 4, the energy bins that provide the largest CNR values are bins 1, 2 and 7. It is evident that the moving phantom data yields much higher CNR for acquisition times greater than 50 ms when compared when the phantom is stationary. This can be explained by the decreasing level of noise in Figure 9, where the noise is the smallest when the phantom is moving at 4 cm/s. Across all energy bins, the noise begins to approach a limit of zero at approximately 50 ms for the stationary and moving phantom data sets and does not significantly decrease thereafter.

In Figures 8d and 8g, the CNR for bins 1, 2 and 7 are overlapped and is likely due to the response of the detector pixels. Since the glass beads are positioned 7 mm apart, they will travel over a different section of the detector. Not to mention, for a fixed acquisition time of 200 ms, the distance in which the contaminants travel will increase with speed, thereby affecting the CNR. Overall, these results demonstrate that imaging a phantom moving at a constant speed in combination with a TDI image reconstruction method is highly advantageous to achieve higher CNR values in comparison to imaging a stationary phantom.

Figure 8: The CNR for the glass beads while stationary (a-c), moving at 1 cm/s (d-f), 2 cm/s (g-i) and 4 cm/s (j-l) for material thicknesses ranging between 2-4 mm. The tube current was 1 mA and no additional filtration was used.

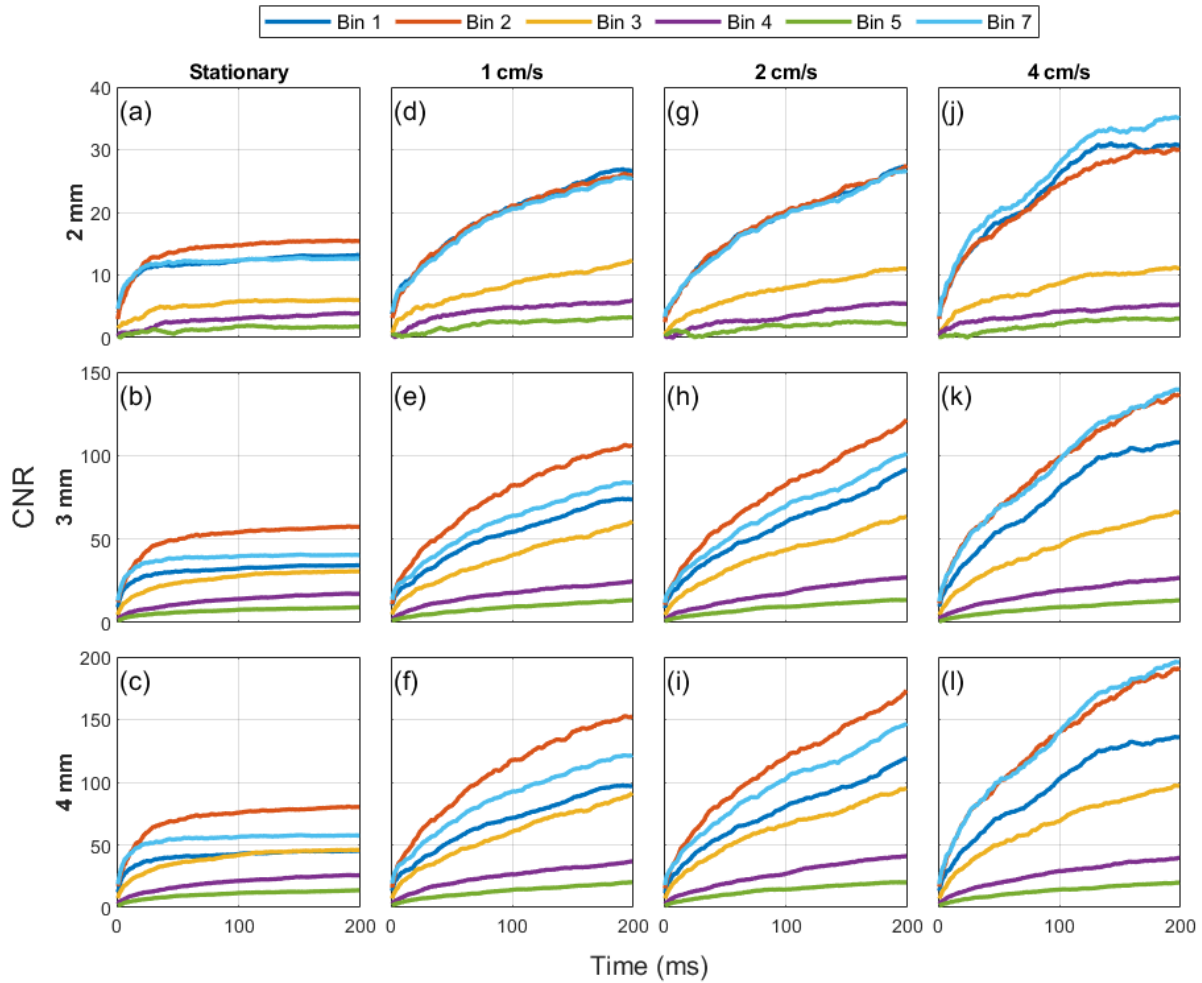

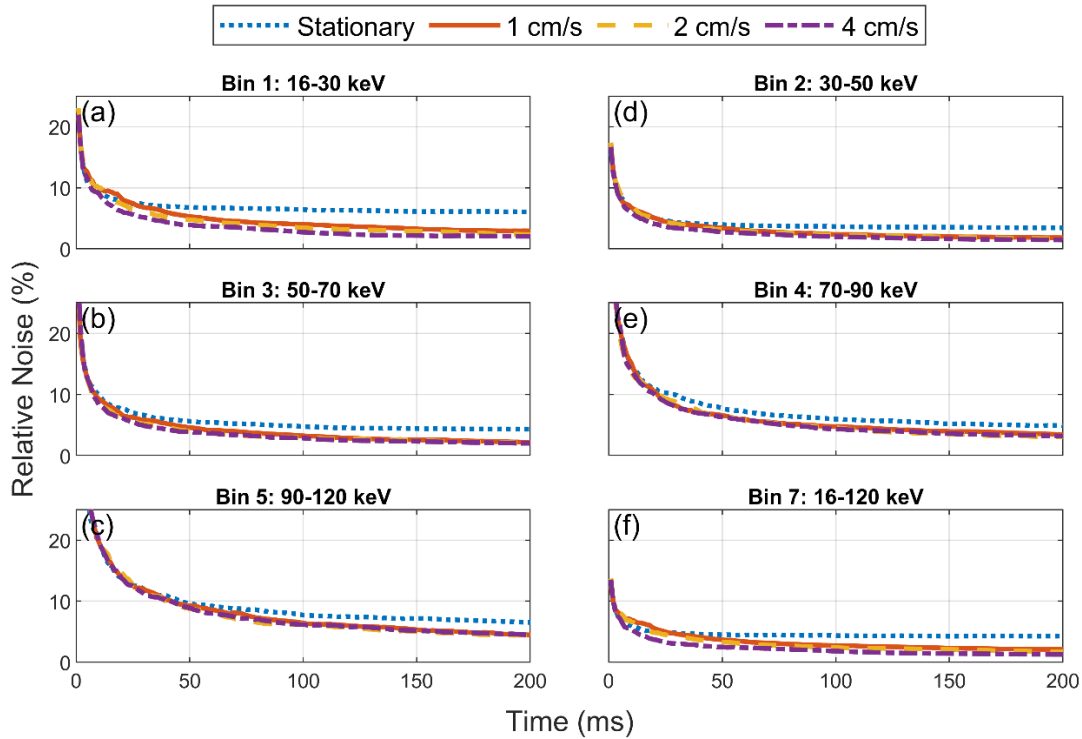

Figure 9: Comparison between the relative noise in the phantom background between the stationary and moving phantom data sets at speeds of 1 cm/s, 2 cm/s and 4 cm/s. The tube current was 1 mA with no external filtration.

3.3 Future work and limitations

Future work aims to increase the phantom speed to be more comparable to belt speeds used in the food industry, which typically ranges between 5-100 cm/s. However, in this work, the sample stage was limited to a maximum speed of 4 cm/s. Other limitations in this study included not correcting for pulse pile up and charge sharing effects, which can degrade the image quality and reduce the CNR. Charge sharing occurs when an incident photon is absorbed near the border of two neighbouring detector pixels and causes the resulting charge cloud to generate two signals in both detector pixels, which can reduce spatial resolution and increase image noise (Willemink et al 2018). Therefore, by accounting for both the pulse pile up and charge sharing effects, further improvements to the CNR and image quality can be achieved.

Previously, our research group showed that a bin width of 10.7 keV and 15.8 keV resulted in the highest CNRs for gadolinium and gold contrast agents for CT imaging, respectively (Richtsmeier et al 2020). Therefore, it is suspected that each type of material will have its own optimal bin width. Alternatively, given that low x-ray energies provide higher image contrast, giving more weight to the low energy bins could significantly improve the CNR for calcium carbonate, polyethylene and Teflon.

For all measurements, the phantom was placed equidistant between the CZT PCD and x-ray tube. However, these distances could be varied to accommodate the size and type of food product being inspected. Therefore, investigating the effect of distance between the phantom, x-ray tube and detector to find the optimal distance for each type of material could further improve the CNR. For a phantom moving at a speed of 4 cm/s, a step size of 4.5 was used in conjunction with a rounding function in Matlab. In this case, it would have been ideal to keep this step size constant and allow the projection images to be shifted by less than a full detector pixel length. Previous work has shown that shifting an image by less than one detector pixel length can be achieved by using a detector shift and iteration method to improve image quality and spatial resolution (Kim et al 2020). This method is especially useful when the object or contaminant in question is smaller than the size of one detector pixel. Following these future improvements, it would be useful to perform image segmentation and classification to identify where the foreign objects are located in the image using clustering methods (O'Connell et al 2019). As well, deep learning techniques has shown to be successful for a wide range of applications in the food industry (Zhou et al 2019; O'Connell et al 2019).

## 4. Conclusions

The emergence of PCDs in the food industry could offer significant advantages over conventional inspection systems that rely on line scan image sensors. The multi-energy capabilities of PCDs allow a wide range of contaminant materials to be better discriminated from the background in which they may be embedded. Based on the results reported in this work, the CNR in bin 2 (30-50 keV) outperformed the other energy in cases in which the stainless-steel, glass and Teflon contaminants were imaged, which is indicative of the advantages of the multi-energy bin capability of PCDs. The system has demonstrated that it is capable of detecting contaminants embedded in a moving phantom, which yielded larger CNR values for image acquisition times larger than 50 ms compared to a stationary phantom. With further improvements to the system, contaminant detection for commercial use is highly plausible.

**Acknowledgements**
This work was funded by the Natural Sciences and Engineering Research Council of Canada (NSERC), Redlen Technologies, and the Canada Research Chair program. The x-ray equipment was funded by the Canada Foundation for Innovation and the British Columbia Knowledge and Development Fund.